\documentstyle[aps,multicol,graphicx]{revtex}
\begin{document}
\title{Large negative velocity gradients in
Burgers turbulence}
\author{A.I.~Chernykh${}^{1,2*}$ and
M.G.~Stepanov${}^{1,2,3\dagger}$}
\address{${}^1$ Institute of Automation and Electrometry,
Novosibirsk 630090, Russia\\ ${}^2$ Novosibirsk State
University, Novosibirsk 630090, Russia\\ ${}^3$
Physics of Complex Systems, Weizmann Institute of
Science, Rehovot 76100, Israel}
\maketitle
\begin{abstract}
  We consider 1D Burgers equation driven by
  large-scale white-in-time random force. The tails of
  the velocity gradients probability distribution
  function (PDF) are analyzed by saddle-point
  approximation in the path integral describing the
  velocity statistics. The structure of the
  saddle-point (instanton), that is velocity field
  configuration realizing the maximum of probability,
  is studied numerically in details. The numerical
  results allow us to find analytical solution for the
  long-time part of the instanton. Its careful
  analysis confirms the result of [Phys. Rev. Lett.
  {\bf 78}, 1452 (1997)] based on short-time
  estimations that the left tail of PDF has the form
  $\ln {\cal P}(u_x) \propto -|u_x|^{3/2}$.

  \vspace{7mm}
\end{abstract}

\begin{multicols}{2}
\section{Introduction}

We consider the random forced Burgers equation
\begin{equation}
  u_t + uu_x - \nu u_{xx} = \phi
\end{equation}
that describes weak 1D acoustic perturbations in the
reference frame moving with the sound velocity
\cite{B74}. The external force $\phi$ in this frame is
generally short-correlated in time, so let us assume
that
\begin{equation}
  \langle \phi(x_1,t_1) \phi(x_2,t_2) \rangle =
    \delta(t_1-t_2) \chi(x_1-x_2).
  \label{phiPC}
\end{equation}
Then the statistics of $\phi$ can be thought Gaussian
and therefore is completely characterized by
(\ref{phiPC}). We are interested in turbulence with a
large value of Reynolds number ${\rm Re} = (\chi(0)
L^4)^{1/3}/\nu$, where $L$ is the characteristic scale
of the stirring force correlator $\chi(x)$. This
problem was intensively studied during the last years
\cite{preP95,preGM96,prlBFKL97,preB97,prlEKMS97,pfGK98}.

The main feature of Burgers turbulence is the
formation of shock waves with large negative velocity
gradient inside and small viscous width of the front.
The positive velocity gradients are decreased by the
dynamics of Burgers equation due to self-advection of
velocity. On the contrary the increasing of negative
gradients could be stopped only by viscosity. The
motion of shock waves leads to a strong intermittency,
the PDF of velocity gradients ${\cal P}(u_x)$ is
strongly non-Gaussian. The one way to describe the
intermittency is to study rare events with large
fluctuations of velocity, that give the main
contribution to the high momenta $\langle u_x^n
\rangle$ or to the PDF tails.

The right tail (positive large $u_x$) of PDF $\ln
{\cal P}(u_x) \propto -u_x^3$ was first found by
Feigel'man \cite{jF82} for the problem of charge
density wave in an impurity potential. Later it was
recovered using operator product expansion
\cite{preP95} (see also \cite{preB97}), instanton
calculus \cite{preGM96}, minimizers approach
\cite{prlEKMS97} and mapping closure \cite{pfGK98}.
The left tail in the inviscid limit seems to be
algebraic, probably ${\cal P}(u_x) \propto
|u_x|^{-7/2}$ \cite{prlEKMS97} (see also
\cite{pfGK98,NBT}). Due to viscosity the very far left
tail is stretched exponential: $\ln {\cal P}(u_x)
\propto -\nu^3 |u_x/\nu|^\beta$. The large negative
gradients exist practically only inside the shock
waves. The maximal value of gradient is proportional
to the square of the velocity jump on the shock wave:
$|u_x|_{\rm max} = (\Delta u)^2/8\nu$. Then roughly
the tail of the shock wave amplitude PDF has the form
$\ln {\cal P}_{\rm shock}(\Delta u) \propto
-\nu^{3-2\beta} (\Delta u)^{2\beta}$. The analysis of
the instanton structure predicts the value $\beta =
3/2$ \cite{prlBFKL97}. This prediction is consistent
with the assumption, that the tails of ${\cal P}_{\rm
shock}(\Delta u)$ should not depend on the viscousity
$\nu$.

We are interested in the statistics of large values of
gradients $u_x \gg u_{\rm rms}/L \sim
(\chi(0)/L^2)^{1/3}$. The velocity field
configurations $u(x,t)$ that make a contribution to
the probability ${\cal P}(a)$ of the equality
$u_x(0,0) = a$ have the gradient greater or equal to
$a$ somewhere. The probability ${\cal P}(a)$ decays
very fast while $a$ increases, i.e. the contribution
of events with gradient greater than $a$ somewhere is
highly suppressed. Then one believes that only some
specific field configurations $u(x,t,a)$ (``optimal
fluctuations'' \cite{ufnL64} or instantons) make
contribution to ${\cal P}(a)$ at large $a > (u_{\rm
rms}/L) \, {\rm Re}$. Under this assumption to
calculate ${\cal P}(a)$ one should find this optimal
field configuration $u(x,t,a)$ and estimate the
probability of its realization.

All instantons of this type are posed at the far tail
of the statistical weight of averaging
$\mu[\phi(x,t)]$. Indeed, to produce large fluctuation
of $u$ the stirring force $\phi$ also should be large,
and the probability of such fluctuation $\phi$ is low.  
The weight $\mu[\phi]$ may not contain a large
parameter, but it should have fast tails, e.g.
exponential ones. Then the concurrence between
statistical weight and the value of calculated
quantity makes the contributing realizations of
$\phi(x,t)$ rather determined. This approach was
introduced by Lifshitz \cite{ufnL64}. Later it was
applyed to determine high order correlation functions
in field theory \cite{jL77} and in the systems of
hydrodynamic type: simultaneous (see, e.g.
\cite{preFKLM96,preGM96,IH}) and non-simultaneous
\cite{IDT} ones.

The paper is organized as follows. In Sec.~\ref{SPA}
we derive the equations for the instanton.
Sec.~\ref{NC} is devoted to the detailed description
of our scheme of numerical calculations. In
Sec.~\ref{VI} we discuss the numerical results and
describe the behavior of the solution of instanton
equations at large times.

\section{Saddle-point approximation}
\label{SPA}

The velocity gradients PDF ${\cal P}(a)$ can be written
as the path integral
\begin{eqnarray}
  && {\cal P}(a) = \big\langle \delta(u_x(0,0)-a)
    \big\rangle_\phi \nonumber \\
  && \,\, {} = \int\! {\cal D}u{\cal D}p \!\!\!
    \int\limits_{-i\infty}^{i\infty} \!\! d{\cal F} \, \exp \Big(
    -{\cal S} + 4\nu^2{\cal F} (u_x(0,0)-a) \Big),
  \label{PDF}
\end{eqnarray}
where the effective action ${\cal S}$ has the form
\cite{Lup,preFKLM96}
\begin{eqnarray}
  {\cal S} = && \frac12 \int\limits_{-\infty}^0 \! dt \int 
    dx_1 dx_2 \, p(x_1,t) \chi(x_1-x_2) p(x_2,t)
    \nonumber \\
  && {} - i\!\int\limits_{-\infty}^0 \! dt \int dx \, p \big(
    u_t + uu_x - \nu u_{xx} \big). \label{S}
\end{eqnarray}
The integration over ${\cal F}$ gives rise to
$\delta(u_x(0,0)-a)$, and the factor $4\nu^2$ was
chosen for our convenience. Note that if retarded
regularization of the path integral (\ref{PDF}) is
used then $\int {\cal D}u{\cal D}p \, \exp(-{\cal S})
= 1$ and we have no normalizing $u$-dependent
denominators in (\ref{PDF}). One can find some
analogies between the appearance of the second field
$p$ and technique that was developed by Keldysh
\cite{jK64} for nonequilibrium dynamics description.

We are interested in the tails of PDF ${\cal P}(a)$,
i.e. the parameter $a$ in the integral (\ref{PDF}) is
large. The asymptotics of ${\cal P}(a)$ at large $|a|
\gg (\chi(0)L)^{2/3}/\nu$ is determined by the
saddle-point configuration of fields $u(x,t)$,
$p(x,t)$ (and also parameter ${\cal F}$), near which
the variation of the integrand is equal to zero
\cite{preFKLM96}. The saddle-point configuration
(sometimes called classical trajectory or instanton)
is governed by the following equations
\begin{eqnarray}
  && u_t + uu_x - \nu u_{xx} = -i\,\chi*p, \\
  && p_t + up_x + \nu p_{xx} = 4i \, \nu^2 {\cal F} \,
    \delta(t)\delta'(x). \label{ie2}
\end{eqnarray}
where $\chi*p$ is the convolution
\begin{equation}
  (\chi*p)(x) = \int dx' \, \chi(x-x') p(x').
\end{equation}
The solution should satisfy boundary conditions
\begin{eqnarray}
  && \lim_{t \rightarrow -\infty} u(x,t) = 0, \quad
      \lim_{t \rightarrow +0} p(x,t) = 0, \nonumber \\
  && \lim_{|x| \rightarrow \infty} u(x,t) = 0, \quad
      \lim_{|x| \rightarrow \infty} p(x,t) = 0.
  \label{BC}
\end{eqnarray}
The value of ${\cal F}$ is tuned in such a way that
the condition $u_x(0,0) = a$ holds. The quantity
${\cal F}$ is a Lagrange multiplier for finding the
extremum of ${\cal S}$ with the condition $u_x(0,0) =
a$.

The equation for $p$ should be solved moving back in
time because of the signs at $p_t$ and $p_{xx}$ in the
instanton equation (\ref{ie2}). The convolution
$-i\chi*p$ is the optimal configuration of external
force $\phi$ that produces large negative gradient.

In what follows we will measure the length in $L$
units, i.e. we set $L = 1$. Rescaling the time $t$ and
fields $u$, $p$ one can exclude the parameter $\nu$
from the instanton equations:
\begin{eqnarray}
  && t = T/2\nu, \quad u = 2\nu U, \quad p = 4i\nu^2 P,
    \quad a = 2\nu A, \\
  && \quad U_T + UU_x - {\textstyle\frac12}
    U_{xx} = \int\! dx' \, \chi(x-x') P(x'),
    \label{IE1} \\
  && \quad P_T + UP_x + {\textstyle\frac12}
    P_{xx} = {\cal F} \, \delta(T)\delta'(x),
    \label{IE2}
\end{eqnarray}
at $T = 0$ one has $U_x(0,0) = A$. The only parameter
in the instanton equations is $A = a/2\nu$. Note that
the steady-state kink solution of Burgers equation
with the negative gradient $a$ is
\begin{equation}
  u = -\sqrt{2\nu|a|} \tanh \Big( \sqrt{|a|/2\nu} \, x
    \Big).
  \label{tanh}
\end{equation}
Thus the physical meaning of $|A|$ is the square of
the ratio of pumping scale $L = 1$ and the kink width
$w_{\rm kink} = 1/\sqrt{|A|}$.

The effective action ${\cal S}_{\rm extr}$ at the
instanton that gives the right exponent:
\begin{equation}
  \ln {\cal P}(a) \simeq -{\cal S}_{\rm extr}(a),
  \label{lnPS}
\end{equation}
is equal to
\begin{eqnarray}
  && {\cal S}_{\rm extr} = -\frac12 \int\limits_{-\infty}^0
    \! dt \int dx_1 dx_2 \, p(x_1,t) \chi(x_1-x_2)
    p(x_2,t) \nonumber \\
  && \,\, {} = 4\nu^3 \!\!\int\limits_{-\infty}^0 \!\! dT \!\int \! dx_1
    dx_2 \, P(x_1,T) \chi(x_1-x_2) P(x_2,T).
  \label{Sextr}
\end{eqnarray}
The freedom of rescaling the fields $u$, $p$ and the
time $t$ with appropriate change of $\nu$ gives us
the following relation:
\begin{equation}
  {\cal S}_{\rm extr}(a) = 8\nu^3 S(a/2\nu) = (2\nu)^3
    S(A),
\end{equation}
with the function $S(A)$ to be determined. One can
prove by straitforward calculation the following
relation between functions ${\cal F}(A)$ and $S(A)$:
\begin{equation}
  {\cal F}(A) = \frac{dS(A)}{dA}.
  \label{FSA}
\end{equation}
The relations of such sort are well-known in classical
mechanics; here $A$ and ${\cal F}$ are conjugate
variables, and saddle-point configuration is the
trajectory of extremal action.

The instanton equations (\ref{IE1},\ref{IE2}) are
Hamiltonian:
\begin{eqnarray}
  U_T(x,T) = -\frac{\delta{\cal H}}{\delta P(x,T)}, 
    \quad P_T(x,T) = \frac{\delta{\cal H}}{\delta
    U(x,T)}, \\
  {\cal H} = \int dx \, P \Big( UU_x -
    {\textstyle\frac12} U_{xx} - {\textstyle\frac12}
    \chi * P \Big). \qquad
\end{eqnarray}
The Hamiltonian ${\cal H}$ is the integral of motion,
i.e. $d{\cal H}/dT = 0$. Since both $U$ and $P$ tend
to zero at $T\rightarrow -\infty$ we have ${\cal H} =
0$. From the instanton equations and the condition
${\cal H} = 0$ we get
\begin{equation}
  S = \frac{{\cal F}A}2 + \frac14 \int dT dx \, P_x
    U^2 = \frac{{\cal F}A}3 + \frac16 \int dT dx \,
    P_x U_x.
\end{equation}
The last term is due to viscousity. At the right tail
it is unimportant, and we have $dS/dA = 3S/A$, i.e. $S
\propto A^3$. At the viscous left tail its
contribution to the action is of the same order as
other terms. If $\ln {\cal P}(a)$ is a powerlike
function: $\ln {\cal P}(a) \propto |a|^\beta$, then
one has $\int dT dx \, P_x U_x = 2(3-\beta)S$.

The high momenta can be calculated by the instanton
method in a following way. Because $a^n {\cal P}(a)$
is a narrow function for large $n$, and only narrow
velocity interval, which position depends on $n$,
contributes to $\langle a^n \rangle$. The position of
this interval is exactly the saddle-point in the
integral $\langle a^n \rangle \propto \int da \, a^n
\exp(-{\cal S}_{\rm extr}(a))$ (see (\ref{lnPS})),
that satisfies the equation
\begin{equation}
  n = a \frac{d{\cal S}_{\rm extr}(a)}{da} = 8\nu^3 A
    \frac{dS(A)}{dA}.
\end{equation}
Combining it with ${\cal F} = n/8\nu^3 A$ we again get
(\ref{FSA}). To get the instanton equations for the
average $\langle a^n \rangle$ one should only
substitute ${\cal F}$ in (\ref{IE2}) for $n/8\nu^3 A$.
Then the instanton equations become the same as in
\cite{prlBFKL97}.

One also should consider fluctuations near the
instanton as a background. The way how the
fluctuations can be taken into account is unknown yet
but their influence to $\ln {\cal P}(a)$ due to their
phase volume is small in comparison with ${\cal
S}_{\rm extr}$ while $a \gg (u_{\rm rms}/L) \, {\rm
Re}$. At smaller gradients the fluctuations
essentially change the answer and we have the
algebraic tail \cite{prlEKMS97}.

\section{Numerical calculations}
\label{NC}

The preliminary calculations that were made in $x$,
$T$ variables have shown that the width of the
instanton equations solution grows with $|T|$ and is
proportional to $|T|^{1/2}$, while its amplitude is
proportional to $|T|^{-1/2}$. To avoid the necessity
of treating simultaneously narrow structure at small
$T$ and wide one at large $T$ we used the following
variables:
\begin{eqnarray}
  && \,\, x = \xi\sqrt{T_0-T}, \quad T = T_0 \left( 1
    - e^{-\tau} \right), \\
  && U = {\tilde U}/\sqrt{T_0-T}, \quad P = {\tilde
    P}/\sqrt{T_0-T},
\end{eqnarray}
where $T_0$ is some constant of the order of unity.
The instanton equations in these variables take the
form
\begin{eqnarray}
  && {\tilde U}_\tau + {\textstyle\frac12} \big(
    \xi{\tilde U} - {\tilde U}_\xi \big)_\xi + {\tilde
    U}{\tilde U}_\xi = \tilde\chi(\tau)*{\tilde P},
    \label{IExt1} \\
  && {\tilde P}_\tau + {\textstyle\frac12} \big(
    \xi{\tilde P} + {\tilde P}_\xi \big)_\xi + {\tilde
    U}{\tilde P}_\xi = {\cal F} \delta(\tau)
    \delta'(\xi)/\sqrt{T_0}, \label{IExt2}
\end{eqnarray}
where $\tilde\chi(\xi,\tau) = (T_0-T)^{3/2}\chi(x)$.
The boundary conditions for ${\tilde U}$, ${\tilde P}$
are analogous to (\ref{BC}).

Let us describe now the general structure of the
numerical scheme that finds the solution of our
boundary problem.

The diffusion terms ${\tilde U}_{\xi\xi}$, ${\tilde
P}_{\xi\xi}$ in instanton equations
(\ref{IExt1},\ref{IExt2}) have opposite signs. If one
considers these equations as two linked Cauchy
problems, then the natural direction of time in
(\ref{IExt1}) is positive, while in (\ref{IExt2}) the
direction is negative. Assume that at a given value of
${\cal F}$ the approximate solution ${\tilde U}_{\rm
old}(\xi,\tau)$ is known. Let us try to make it closer
to the true solution of the problem. For this purpose
let us solve the Cauchy problem for (\ref{IExt2})
starting from $\tau = +0$ and moving up to large
enough $\tau_{\rm min} < 0$. Then using ${\tilde P}$
that we have got in the previous step we solve the
Cauchy problem for (\ref{IExt1}) moving from $\tau =
\tau_{\rm min}$ up to $\tau = 0$. As a result we get
the new values ${\tilde U}_{\rm new}(\xi,\tau)$.
Further we will use the sign $f$ for the mapping
${\tilde U}_{\rm old} \rightarrow {\tilde U}_{\rm
new}$. The stationary point ${\tilde U}$ of the
mapping $f$ and the corresponding function ${\tilde
P}$ are the desired solution of
(\ref{IExt1},\ref{IExt2}).

The numerical experiments have shown that iterations
\begin{equation}
  {\tilde U}^{(i+1)} = f \big[ {\tilde U}^{(i)} \big],
    \quad {\tilde U}^{(0)} \equiv 0 \label{SI}
\end{equation}
converge if ${\cal F} > {\cal F}_* \simeq -0.96$. 
While ${\cal F} < {\cal F}_*$ the simple iterations
(\ref{SI}) are divergent.

\begin{figure}
  \begin{center}
    \includegraphics[width=3in]{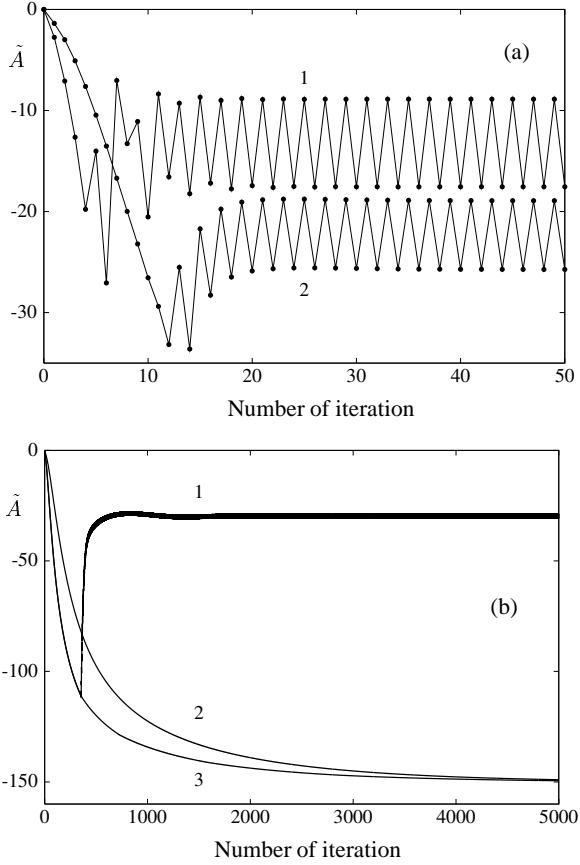}
  \end{center}
  \caption{The gradient ${\tilde A}$ as a
    function of number of iteration. At
    Fig.~\protect\ref{fig1}(a) the dots corresponding
    to one run are joined by a line for convenience. 
    The curves are: (a) simple iterations according to
    (\protect\ref{SI}) (curve 1) and according to
    (\protect\ref{Fc}) with $c = 0.1$ (curve 2); (b)
    $c = 0.1$ (curve 1, its thickness is determined by
    the amplitude of serrated oscillations), $c =
    0.05$ (curve 2), and adaptive decreasing of $c$
    during the run, the initial value is $c = 0.1$,
    $c$ was multiplied by $0.9$ four times (curve 3);
    ${\cal F} = -2$.}
  \label{fig1}
\end{figure}

Curve 1 at the Fig.~\ref{fig1}(a) shows how the value
of the gradient ${\tilde A} = \partial {\tilde U}
(\xi,\tau)/\partial \xi|_{\xi = 0,\tau = 0}$ depends
on the number of iteration for ${\cal F} = -2$. It can
be seen that the stationary point of $f$ is unstable,
but the mapping $f \big[ f \big[ {\tilde U} \big]
\big]$ has two stable stationary points. The stability
properties of the iterations are determined by the
spectrum of the linearization ${\hat K}$ of $f$ near a
stationary point:
\begin{equation}
  f \big[ {\tilde U}+V \big] = {\tilde U} + {\hat K}
    V + \dots
\end{equation}
The iteration process is convergent if the modulus of
all eigenvalues of the linear part ${\hat K}$ is less
than 1. The period doubling indicates that while
${\cal F}$ passes through ${\cal F}_*$ one of ${\hat
K}$'s eigenvalues passes through $-1$
\cite{PeriodDoubling}. Let us denote this eigenvalue
as $\lambda$.

This knowledge allows us to construct the new mapping
with stable stationary point that coincides with one
of the mapping $f$. Let us proceed the following
iterations:
\begin{equation}
  {\tilde U}^{(i+1)} = f_c \big[ {\tilde U}^{(i)} \big]
    \equiv c\,f \big[ {\tilde U}^{(i)}\big] + (1-c)
    {\tilde  U}^{(i)},
  \label{Fc}
\end{equation}
where $c$ is some constant. It is easy to check that
the stationary points of the mappings $f$ and $f_c$ do
coincide. The linear part of $f_c$ is equal to ${\hat
K}_c = c {\hat K} +(1-c){\hat 1}$, its eigenvalue
corresponding to unstable $\lambda$ is $\lambda_c =
c\lambda + (1-c)$. If we take the value of $c$ inside
the interval $0 < c < 2/(1-\lambda) < 1$, then
$|\lambda_c|<1$, and the iterations (\ref{Fc}) are
convergent.

Since we don't know $\lambda$ a priori, the value of
$c$ that provides the convergence of iterations, was
determined from experiment. Fig.~\ref{fig1}(a)
illustrates the influence of the decreasing of $c$ on
the dependence of ${\tilde A}$ vs. number of
iteration. In the final version of the computer code
the value of $c$ was changed in an adaptive way: each
time the value of $|{\tilde A}|$ was decreased after an
iteration, $c$ was multiplied by $0.9$. One can
compare from the Fig.~\ref{fig1}(b) the iterations run
for $c = 0.1$, $c = 0.05$ and for adaptive decreasing
of $c$, all three for ${\cal F} = -2$. The initial and
final values of $c$ in the case of adaptive change
were equal to $0.1$ and $0.1 \cdot 0.9^4 =0.05905...$,
respectively. It is clear that for the two last cases
the iterations converge to the same solution.

\subsection{Grid parameters}

The solution of Cauchy problem is found numerically
using the method of finite differencies. The grid
covers the rectangular domain $0 < \xi < \xi_{\rm
max}$, $\tau_{\rm min} < \tau < 0$. In numerical
calculations some of the boundary conditions (\ref{BC})
that in principle are posed at the infinity were
considered as they are posed at (large enough)
$\xi_{\rm max}$ and $\tau_{\rm min}$. Typical values
used: $\xi_{\rm max} = 10$ and $\tau_{\rm min} = -30$.
The grid had uniform mesh intervals in $\xi$, typical
number of grid sites along $\xi$ axe was eqial to
$1024$.

The first calculations had shown that the solution
changes rapidly in the vicinity of $\tau = 0$, while
at large $\tau$ it varies slowly. Because of this we
used nonuniform grid for variable $\tau$. The time
step was smaller inside the interval $\tau_1 < \tau <
0$, typical value used: $\tau_1 = -0.4$. The number of
grid sites inside this interval varied from $2000$ to
$4000$. The computer power we had limited the number
of all time steps in the grid by $5000$. During all
calculations we used $T_0=1$.

\subsection{Cauchy problem for $\tilde P$}

The equation (\ref{IExt2}) should be solved backward
in time. The source term in right hand side of the
equation (\ref{IExt2}) provides us the initial
condition ${\tilde P}(\tau = -0) \propto
\delta'(\xi)$. This means that at small times the
field ${\tilde P}$ is localized in a very narrow
interval centered at $\xi = 0$. Such initial condition
can not be accurately discretized, so we used another 
way to represent ${\tilde P}$ at small times.

For small $\tau$ the field ${\tilde P}$ is very
narrow. At its support we can approximate the velocity
${\tilde U}$ by a linear profile: ${\tilde U} =
{\tilde A}(\tau) \xi$. The evolution of ${\tilde P}$
in such a velocity field is described by the
derivative of Gaussian contour:
\begin{eqnarray}\label{TNZprime}
  && \tilde P(\xi,\tau) = {\cal F} \, \frac{\xi P_{\rm
    amp}(\tau)} {\sqrt{2\pi T_0 D^3(\tau)}} \exp
    \left( -\frac{\xi^2}{2D(\tau)} \right), \\
  && P_{\rm amp}(-0) = 1, \quad D(-0) = 0. \nonumber
    \\
  && D(\tau) = \int\limits_\tau^0 d\tau' \, \exp \left(
    \int\limits_{\tau'}^\tau d\tau'' \, (2\tilde A(\tau'') +
    1) \right), \\
  && P_{\rm amp}(\tau) = \exp \left( -\int\limits_\tau^0
    d\tau' ( 2\tilde A(\tau')+1/2 ) \right).
\end{eqnarray}
We use such a representation for ${\tilde P}$ for
$\tau_0 < \tau < 0$. The value of $\tau_0$ is chosen
in such a way, that the velocity field ${\tilde U}$ is
still linear at the width of ${\tilde P}$. From the
other hand, ${\tilde P}$ at $\tau = \tau_0$ already
becomes wide in comparison with mesh interval
$\Delta\xi$. Typical value of $\tau_0$ used: $\tau_0 =
-1/1500$. For times $\tau < \tau_0$ the solution is
found by fully implicit scheme
\begin{eqnarray}
  && \frac{{\tilde P}_i^{n+1} - {\tilde P}_i^n}
    {\Delta \tau} + \frac12 {\tilde P}_i^{n+1} +
    D_i^{n+1} \frac{{\tilde P}_{i+1}^{n+1} - 2 {\tilde
    P}_i^{n+1} + {\tilde P}_{i-1}^{n+1}} {\Delta
    \xi^2} \nonumber \\
  && \,\, {} +
    r_{+,i}^{n+1} \frac{{\tilde P}_{i+1}^{n+1} -
    {\tilde P}_i^{n+1}} {\Delta \xi} + r_{-,i}^{n+1}
    \frac{{\tilde P}_{i}^{n+1} - {\tilde
    P}_{i-1}^{n+1}} {\Delta \xi} = 0, \label{RSP}
\end{eqnarray}
where $r_{\pm,i}^n$, $D_i^n$ are equal to
\begin{eqnarray}
  && r_{\pm,i}^n = 0.5 \big( r(\xi_i,\tau_n) \pm
    |r(\xi_i,\tau_n)| \big), \nonumber \\
  && D_i^n = \frac{1}{1+0.5|r(\xi_i,\tau_n)|\Delta\xi}.
  \label{SPCF}
\end{eqnarray}
The function $r(\xi,\tau)$ is expressed via velocity
field: $r(\xi,\tau) = 0.5\,\xi + {\tilde U}
(\xi,\tau)$.

Here $\Delta\xi>0$, $\Delta\tau < 0$ --- mesh
intervals, and $\xi_i$, $\tau_n$ --- site coordinates.
The numerical scheme used is monotonous and stable, it
is of the first order of accuracy in $\Delta\tau$ and
of the second order in $\Delta\xi$ \cite{SamGul}.

\subsection{Cauchy problem for $\tilde U$}

At this stage we use the initial condition ${\tilde
U}(\tau = \tau_{\rm min}) \equiv 0$. The viscousity,
source and self-advection terms are treated by
splitting technique \cite{Rasshchep}. At each time
step it is first calculated the change of ${\tilde U}$
due to the source, then --- to the viscousity, and at
the last --- to the nonlinearity.

From calculated already grid layer ${\tilde U}^n =
{\tilde U}(\tau = \tau_n)$ the next layer ${\tilde
U}^{n+1}$ is found in a following order: first, the
equation ${\tilde U}_\tau = {\tilde\chi}(\tau)*{\tilde
P}$, is solved:
\begin{equation}\label{RSou}
  \frac{\tilde U_i^{n+1}-\tilde U_i^n}{\Delta\tau}
    = (\tilde\chi(\tau)*{\tilde P})_n.
\end{equation}
The convolution ${\tilde\chi} * {\tilde P}$ is
calculated as the result of inverse fast Fourier
transform (FFT) acting on the product of ${\tilde
\chi}$'s and ${\tilde P}$'s FFT images. The external
force correlator during all calculations was equal to
$\chi(x) = (1-x^2) e^{-x^2/2} = -d^2 e^{-x^2/2}/dx^2$.
The numbers ${\tilde U}_i^{n+1}$ that are found in
such a way {\it are not\/} a final solution for the
layer ${\tilde U}^{n+1}$, since only the source term
has been taken into account yet. We use them as an
input at the next step, we will denote them as
${\tilde U}^n_i$ (note that they {\it do not\/}
coincide with ${\tilde U}^n_i$ in (\ref{RSou})).

Next, the viscousity and linear part of advection are
taken into account, according to the equation ${\tilde
U}_\tau + {\textstyle\frac12} \big( \xi{\tilde
U}-\tilde U\big)_\xi = 0$. The fully implicit scheme
was used analogously to (\ref{RSP}):
\begin{eqnarray}
  && \frac{{\tilde U}_i^{n+1} - {\tilde U}_i^n}{\Delta
    \tau} + \frac12 {\tilde U}_i^{n+1} - D_i
    \frac{{\tilde U}_{i+1}^{n+1} - 2 {\tilde
    U}_i^{n+1} + {\tilde U}_{i-1}^{n+1}} {\Delta
    \xi^2} \nonumber \\
  && \qquad\quad {} + \frac12 \frac{{\tilde U}_{i}^{n+1} -
    {\tilde U}_{i-1}^{n+1}} {\Delta \xi} = 0,
  \label{RSU}
\end{eqnarray}
here $D_i = 1/(1 + 0.25 \xi_i \Delta\xi)$. Again, the
numbers ${\tilde U}_i^{n+1}$ do not form a final
solution, and we send them to the next step with a
${\tilde U}_i^n$ notation.

The nonlinear part of the equation ${\tilde U}_\tau +
{\tilde U}{\tilde U}_\xi = 0$ was solved by explicit
conservative scheme \cite{Musher}:
\begin{equation}\label{RSHopf}
  \frac{{\tilde U}_i^{n+1} - {\tilde U}_i^n}{\Delta
    \tau} + \frac{{\tilde U}_{i+1}^n + {\tilde U}_i^n 
    + {\tilde U}_{i-1}^n}3 \; \frac{{\tilde U}_{i+1}^n
    - {\tilde U}_{i-1}^n} {2\Delta \xi} = 0,
\end{equation}
that finally gives us the next layer of the velocity
field ${\tilde U}^{n+1}$. This scheme is of the first
order of accuracy in $\Delta \tau$ and of the second
one in $\Delta \xi$.

\section{Viscous instanton}
\label{VI}

In this section we represent the results of our
calculations, that show the structure of the instanton
and its change with $|{\cal F}|$.
The minimal value of ${\cal F}$ at which the reliable
results in numerics were obtained is ${\cal F} = -2$.

\begin{figure}
  \begin{center}
    \includegraphics[width=3in]{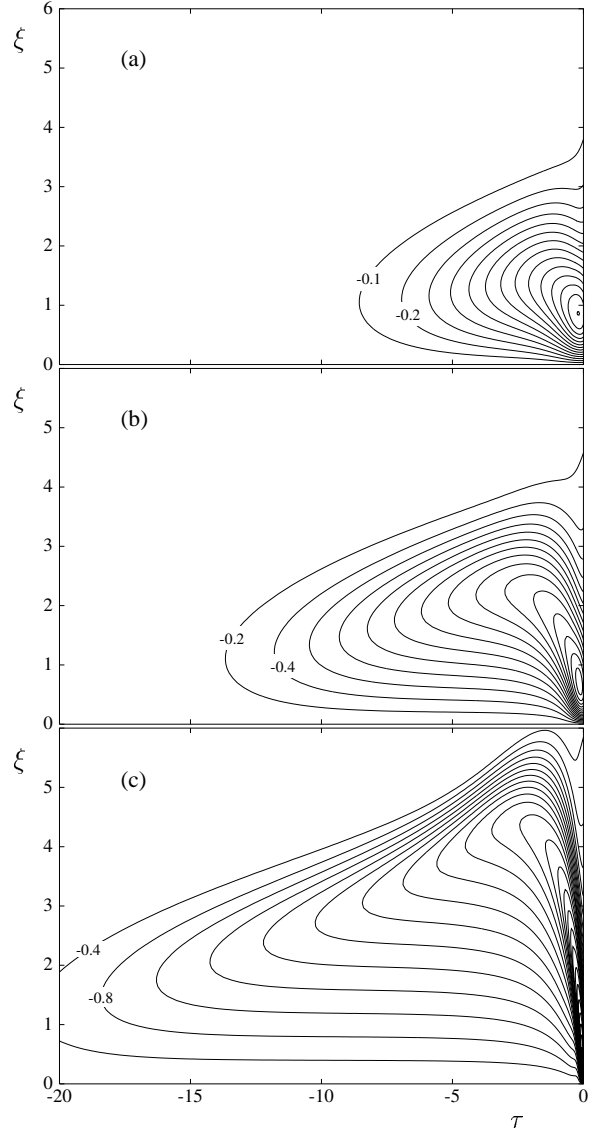}
  \end{center}
  \caption{The level curves if ${\tilde U}(\xi,\tau)$
    for ${\cal F} = -0.9$ (a), ${\cal F} = -1.1$ (b),
    and ${\cal F} = -2$ (c). The values of levels can
    be calculated from the two given levels according
    to arithmetic progression law.}
  \label{fig2}
\end{figure}

The general features of the instanton structure change
with ${\cal F}$ can be obtained from Fig.~\ref{fig2}
that shows the level curves of ${\tilde
U}(\xi,\tau)$ for three values of ${\cal F}$. Since
$\tilde U$ and $\tilde P$ are the odd functions of
$\xi$, we draw only the region where $\xi > 0$. The
calculations were done in a rectangular $0 < \xi <
10$, $-30 < \tau < 0$, whose dimensions are a bit
larger than it is shown at the figure.

One can see that the instanton life-time and the
maximal value of $|{\tilde U}|$ rapidly increase
with the growth of $|{\cal F}|$. The growth leads
to the deformation of level curves near $\tau = 0$
because of the influence of nonlinearity, that is
weak at ${\cal F}=-0.9$ (Fig.~\ref{fig2}(a)) and very
strong at ${\cal F}=-2.0$ (Fig.~\ref{fig2}(c)).

\subsection{Structure}

Detailed analysis of the instanton solution based on
the results of numerical calculations allows us to
distinguish five different regimes in the instanton
time evolution. Below we discuss them consequently
from $t = 0$ to $t = -\infty$.

\begin{figure}
  \begin{center}
    \includegraphics[width=3in]{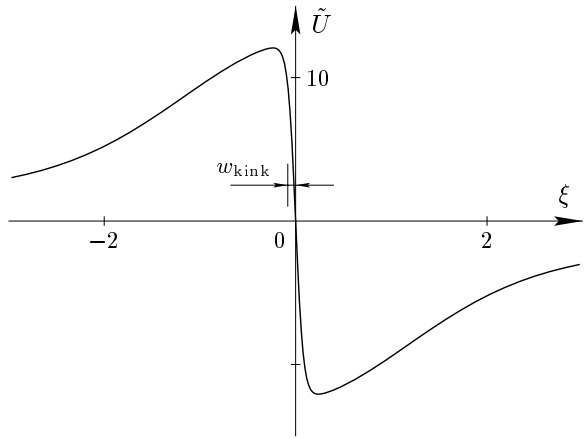}
  \end{center}
  \caption{The kink in the velocity field ${\tilde
    U}(\xi,\tau)$ at $\tau = 0$; ${\cal F} = -2$.}
  \label{fig3}
\end{figure}

The first regime consist in the viscous smearing of
the field $p$ up to the scale of the kink width
$w_{\rm kink} = \sqrt{2\nu/|a|} = 1/\sqrt{|A|}$ (see
(\ref{tanh}), Fig.~\ref{fig3}). Since the viscousity
plays the crucial role in this regime, we will also
use dimensional variables. At $t = -0$ we have $p(x,t)
\propto \delta'(x)$, and the width of the kink in the
velocity profile is equal to $w_{\rm kink} =
\sqrt{2\nu/|a|}$ (see (\ref{tanh})). Since $p$ is very
narrow, the viscousity dominates the evolution. The
width of $p$ obeys the diffusion law and equals to
$\sqrt{2\nu|t|}$. These two widths become comparable
at time $t = -1/|a|$, or $T = -1/|A|$. This means that
during the whole time of smearing of $p$ by viscousity
the width of the kink is of the order of $w_{\rm
kink}$. Indeed, if the shape of the velocity profile
deviates from steady-state kink solution (\ref{tanh}),
then the change of the kink width during this time
period would be of the order of $ut \sim
\sqrt{2\nu|a|}/|a| = \sqrt{2\nu/|a|} = w_{\rm kink}$.
At this regime the source term $\chi * P$ in instanton
equation (\ref{IE1}) is unimportant. When the width of
$p$ becomes of the order of $w_{\rm kink}$ the rate of
expansion of $p$ due to the velocity gradient becomes
comparable with the rate due to viscousity (such a
balance determines the width of the kink).

The next (second) regime was exhaustively studied in
\cite{prlBFKL97}. It consists in dilation of fields
$U$, $P$ up to the pump scale $L = 1$. The fields are
advected by velocity $U$, and considering evolution
back in time they are expanded by it since $U_x|_{x =
0} < 0$. The time needed for the expansion is equal to
$T_* \sim L/U \sim 1/\sqrt{|A|}$.

\begin{figure}
  \begin{center}
    \includegraphics[width=3in]{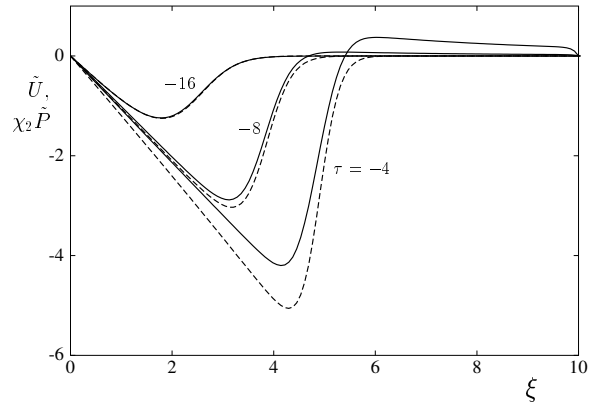}
  \end{center}
  \caption{Comparison of ${\tilde U}(\xi,\tau)$ (solid
    curves) and $\chi_2 {\tilde P}(\xi,\tau)$ for
    three values of $\tau$; ${\cal F} = -2$.}
  \label{fig4}
\end{figure}

During the 3rd--5th regimes the width of $U$ and $P$
field is much greater than $L = 1$. Then it is naturally
to substitute $\chi(x)$ by $-\chi_2\delta''(x)$. The
instanton equations take the following simple form:
\begin{equation}
  U_T + UU_x + {\textstyle\frac12} V_{xx} = 0, \quad
  V_T + UV_x + {\textstyle\frac12} U_{xx} = 0,
  \label{EqUV}
\end{equation}
where $V = 2\chi_2 P - U$, $\chi_2 = -\frac12 \int dx
\, x^2 \chi(x)$. While moving to large negative time
numerical solution has a tendency to fall in $U = V$
(see Fig.~\ref{fig4}, curves $\tau=-4$ and $\tau=-8$).
Then the equations for $U$, $V$ reduce to the Burgers
equations with the evolution back in time.

\begin{figure}
  \begin{center}
    \includegraphics[width=3in]{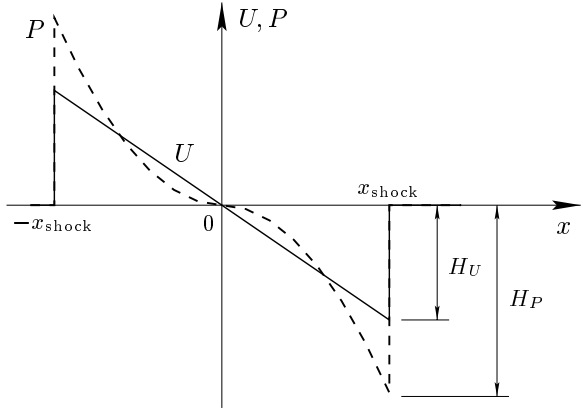}
  \end{center}
  \caption{Schematic representation of $U$ and $P$ for
    the 3rd regime, ${\cal F} \rightarrow -\infty$.}
  \label{fig5}
\end{figure}

Such a substitution is not always possible. During
such an evolution the shock waves occure (see, e.g.,
the level curves at Fig.~\ref{fig2}(c) near $\tau =
-5$, Fig.~\ref{fig4}, curves 1 and 2). For transition
to equations (\ref{EqUV}) to be valid the width of
these shock waves should be larger than $L = 1$.
Otherwise the substitution of $\chi$ by $\delta''$ is
not valid. The width of the shock wave in
dimensionless variables is greater than $1$ only if
its height is smaller than $1$. However, right after
the 2nd regime when the width of $U$ and $P$ fields
becomes greater than the pumping force correlation
length $L = 1$ the amplitude of the velocity field $U$
is of the order of $\sqrt{|A|} \gg 1$. The amplitude
of $U$ becames small only at very large times (it is
shown below that such a crossover happens at $T \sim
-\sqrt{|A|}$). It means that there is an intermediate
regime that goes after the 2nd one, where the
substitution of $\chi$ by $\delta''$ is inapplicable.
In this (third) regime the fields $U$ and $P$ are
smooth functions in the interval wider than $1$. At
the ends of this interval they contain shock waves ---
the value of $U$ and $P$ rapidly goes to zero (as it
is shown schematically in Fig.~\ref{fig5}). We will
use the word ``shock'' for these structures at the
ends of the interval, while for narrow structure in
the velocity field $U$ near $x = 0$, $T = 0$ we will
use the word ``kink''.

Now we will consider the structure of the shocks in
detail. Let us denote the height of the shocks in
$U(x,T)$ and $P(x,T)$ fields by $H_U(T)$ and $H_P(T)$,
respectively. Their position we will denote as $x_{\rm
shock}(T)$ ($x_{\rm shock} > 0$, shocks are posed at
$x = \pm x_{\rm shock}$).

\begin{figure}
  \begin{center}
    \includegraphics[width=3in]{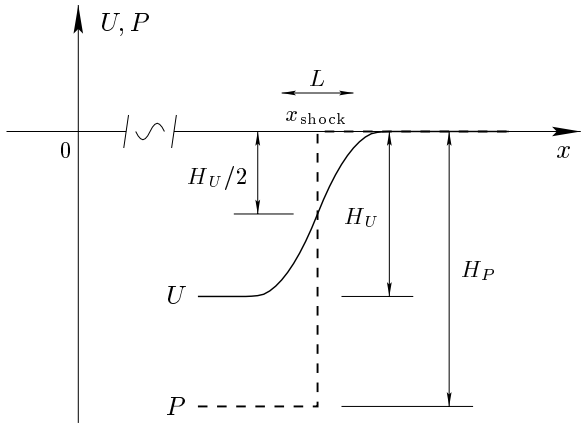}
  \end{center}
  \caption{Shock structure in the 3rd regime, ${\cal
    F} \rightarrow -\infty$.}
  \label{fig6}
\end{figure}

In the third regime the width of the shocks in the
field $P$ is determined by the competition of
squeezing them by the velocity $U$ and spreading by
viscousity (see Fig.~\ref{fig6}). Since we are in a
strongly nonlinear regime the viscousity is weak, and
the width of $P$'s shocks is very small. The shocks in
$P$ are stationed at the center of $U$'s shocks, i.e.
in almost linear velocity profile with the gradient
$U_x \sim H_U$. The width of $P$'s shocks can be
estimated as $1/\sqrt{H_U} \ll L$. Then the good
approximation for $P(x,T)$ near the shock $x \simeq
x_{\rm shock}$ is $P(x,T) \simeq -H_P(T) \theta(x_{\rm
shock}(T)-x)$, where $\theta(x)$ is step-function.
During the evolution forward in time the shocks in $U$
do not break down because of source term $\chi * P$.
The source prevents the destructive effect of
advection term $UU_x$. The shocks in $P$ should carry
the $U$'s shocks of the height $H_U$, i.e. of strength
$UU_x \sim H_U^2$. Thus we should have $H_P \propto
H_U^2$ in this regime. Now we show it more carefully.

Let us write the instanton equation (\ref{IE1}) in the
reference frame of the shock near the point $x =
x_{\rm shock}$ (see Fig.~\ref{fig6}).  We have two
contributions to time derivative $U_T$: from the
growth of $H_U$ in time (of order $H_U/T$) and from
the motion of the shock (of order $H_U^2$). Neglecting
the first one we write the following equation for the
velocity $U(x,T)$:
\begin{equation}
  \frac12 \big( U(x)-U(x_{\rm shock}) \big)^2_x = -H_P
    X'(x-x_{\rm shock}),
\end{equation}
where the new function $X(x)$ is determined by the
equation $\chi(x) = -X''(x)$ with the condition $X
\rightarrow 0$ with $x \rightarrow \pm \infty$.
Integrating this equation once we obtain
\begin{equation}
  \big( U(x)-U(x_{\rm shock}) \big)^2 = 2 H_P \big( X(0)
    - X(x-x_{\rm shock}) \big).
\end{equation}
Since $U(x_{\rm shock}) = -H_U/2$ we have
$H_P = H_U^2/8X(0)$.

The next step consists in finding the solution of
(\ref{IE1},\ref{IE2}) between the shocks posed at $x =
\pm x_{\rm shock}$ considering the fields $U$, $P$ as
smooth ones and using the boundary condition
\begin{equation}
  P(\pm x_{\rm shock} \mp 0,T) = \mp U^2(x_{\rm shock}-0,T)/8X(0).
    \label{BCUP}
\end{equation}
The instanton equations (\ref{IE1},\ref{IE2}) take the
form
\begin{equation}
  U_T + UU_x = 0, \quad P_T + UP_x = 0.
  \label{EqUP}
\end{equation}
Here we approximate the shocks as jump
discontinuities, and the condition (\ref{BCUP})
relates the heights of the jumps (see
Fig.~\ref{fig5}). Here the diffusion terms and the
term $\chi * P$ are omitted. One can check that they
are negligible since the characteristic $x$-scale of
the solution is large enough.

The equations (\ref{EqUP}) can be integrated by
characteristics (or Lagrangian trajectories).
The velocity of the shocks is equal to $\pm H_U/2$,
i.e. all the trajectories disappear at the shocks
(if we consider the evolution back in time).
The value of $U$ (or $P$) is conserved in time
if we follow the Lagrangian trajectory. This
means that the relation $P = U|U|/8X(0)$ holds
everywhere between the shocks.

\begin{figure}
  \begin{center}
    \includegraphics[width=3in]{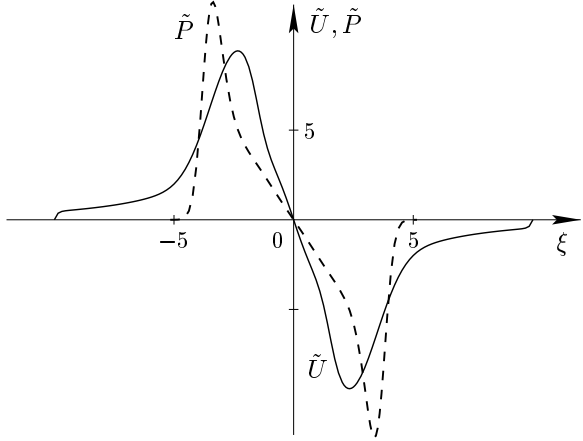}
  \end{center}
  \caption{${\tilde U}(\xi)$ (solid curve) and $\chi_2
    {\tilde P}(\xi)$ (dashed curve) profiles at $\tau
    = -0.4$. ${\cal F} = -2$.}
  \label{fig7}
\end{figure}

Due to self-advection the velocity field $U$ becames
more and more linear as a function of $x$ while $|T|$
increases.  This happens at the border between the 2nd
and the 3rd regimes. In the third regime we can take
$U(x,T) = x/T$ between the shocks. The field $P$ is
equal to $P(x,T) = -x|x|/8X(0)T^2$. The velocity $U$
simply squeezes or expands the field $P$ without
changing its shape. Since $H_P \sim H_U^2$, the field
$P$ should have the same scaling as $U^2$, i.e. $P
\propto x^2$. The concave form of $P$ (as one can see
${\cal F} = -2$ is not yet good enough for a clear
picture) is shown in Fig.~\ref{fig7}.

Let us determine the time dependence of $x_{\rm shock}$.
We have
\begin{equation}
  \frac{d x_{\rm shock}}{dT} = -\frac12 H_U(T) =
    -\frac{x_{\rm shock}}{2T}.
\end{equation}
Solving this equation we get $x_{\rm shock}(T) = B
\sqrt{-T}$. Since $x_{\rm shock} \sim 1$ at $|T| \sim
1/\sqrt{|A|}$, we get $B \sim |A|^{1/4}$. The shocks heights
are equal to $H_U(T) = B/\sqrt{-T}$, $H_P(T) = B^2/8
X(0)|T|$. The width of the shocks in $P$ field is
of the order of $1/\sqrt{H_U}$. The shocks in velocity
field $U$ have the width of the order of $1$, since $U$
is pumped by $\chi * P$.

At large time $T \sim -\sqrt{|A|}$ the height $H_U$
(and consequently the width of the shocks in $P$)
becames of the order of $1$. This indicates the end
of the third regime and the beginning of the fourth. 
Going further back in time we
finally enter the domain of validity of the equations
(\ref{EqUV}). The solution falls into $U = V$. Again,
the solution has two shocks between which it is a
smooth function. The shock position satisfies
$x_{\rm shock} \propto \sqrt{-T}$, between the shocks
$U = V = x/T$ holds. This regime exactly
corresponds to self-similar solution $u(x,t) =
\theta(t - C x^2) \, x/t$ of
inviscid Burgers equation $u_t + uu_x - 0 \cdot
u_{xx} = 0$ (here $C$ is a parameter).

At the far tail of the instanton due to viscous
dissipation the solution $U = V$ transformes to the
derivative of Gaussian contour --- the self-similar
solution of the diffusion equation (see
Fig.~\ref{fig4}, $\tau = -16$). During the fifth
regime the advection term $UU_x$ becomes irrelevant.
In $\xi$, $\tau$ variables the solution tends to
${\tilde U} \propto \xi \exp (\tau/2 - \xi^2/2)$, that
was observed in numerical calculations.

During the 4th and the 5th regimes the amplitude of
the velocity field $U$ is less than unity. It means
that the amplitude of $U$ is lower than the order of
the typical statistical fluctuations, and the
saddle-point approximation is meaningless there. The
typical events that demonstrate large negative
gradients start from some velocity configuration
$U(x)$ with the amplitude of the order of unity and
are governed by the 3rd regime first. The action on
these events and their further evolution almost do not
depend on the initial velocity field $U(x)$, and the
dependence of ${\cal P}(a)$ on $a$ remains unaltered
by the averaging over all possible configurations
$U(x)$. We considered the 4th and the 5th regimes
since they are the parts of the whole solution of our
nonlinear boundary problem.

\begin{figure}
  \begin{center}
    \includegraphics[width=3in]{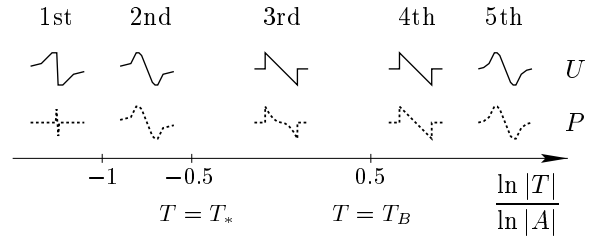}
  \end{center}
  \caption{Schematic representation of the instanton
    time structure.}
  \label{fig8}
\end{figure}

Let us now run the whole evolution forward in time. At
the beginning (5th regime) the field $U$ is pumped by
very wide $P$. The pumping force is proportional to
$P_{xx}$. During the 4th regime the source is
localised at shocks in $P$ that leads to a formation
of shocks in $U$. The $U$'s shocks want to break down
because of self-advection, but the source term $\chi *
P$ keeps them going. When the growing height of $P$
becomes larger than unity the shocks in $P$ become
narrow. The balance between the terms $UU_x$ and $\chi
* P$ in (\ref{IE1}) changes a little, that results in
the change of the form of $U$'s shocks --- it is
determined by the shape of $\chi(x)$ now. The distance
between the $P$'s shocks decreases in time and
eventually it becomes comparable with unity. After
this $P$ becomes even more narrow and the efficiency
of the source term begins to fall down. The
self-advection of the velocity destroyes the shocks
and leads to a formation of the kink at $x = 0$, while
$P$ transformes to $\delta'(x)$. The kink shape at
${\cal F} = -2$ is shown in Fig.~\ref{fig3}.
Schematically time evolution of the instanton is
illustrated in Fig.~\ref{fig8}.

\subsection{Action}

\begin{figure}
  \begin{center}
    \includegraphics[width=3in]{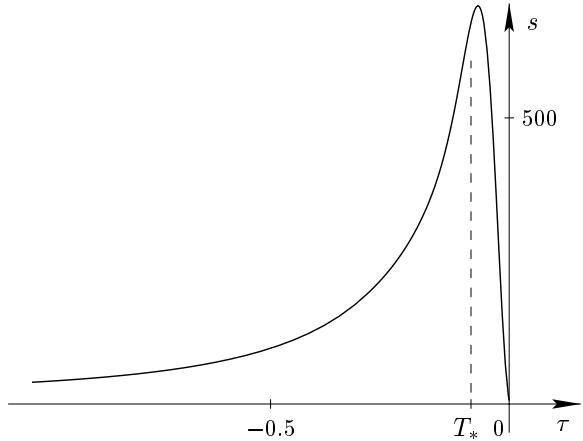}
  \end{center}
  \caption{The action density $s(\tau)$ as a function
    of ``time'' $\tau$. ${\cal F} = -2$.}
  \label{fig9}
\end{figure}

One can present the action $S(A)$ in the form of $S =
\int_{-\infty}^0 dT \, s(T)$ like in expression
(\ref{Sextr}). For ${\cal F} = -2$ the action density
$s(T)$ that was obtained from numerical calculations
is shown in Fig.~\ref{fig9}. While $T < T_* =
-1/\sqrt{|A|}$ the convolution $\chi*p$ is localized
at shocks, so $s(T) \sim H_P^2(T) \sim B^4/T^2$. The
maximum of $s(T)$ is posed at $T \sim T_*$. Further
increasing $T$ leads to the decreasing of the density
$s(T)$ because $P(x)$ becomes more and more narrow
without an adequate growth of its amplitude. This
region of small time $T > T_*$ was studied in
\cite{prlBFKL97}. It was shown that the contribution
$S_{T > T_*}$ to the extremal action from this
interval is of order $|A|^{3/2}$, and the main
contribution to the action from it comes from the
region $T \sim T_*$ --- the border between the 2nd and
the 3rd regimes. Exactly these two regimes determine
the optimal configuration of noise providing the event
with large negative gradient.

The contribution of region of time $-\sqrt{|A|} = T_B
< T < T_* = -1/\sqrt{|A|}$ (3rd regime) to the extremal
action $S(A)$ can be estimated as
\begin{equation}
  S_{T < T_*} \sim \int\limits_{T_B}^{T_*} dT \, H_P^2(T)
    \sim -B^4/T_* \sim |A|^{3/2}. \label{SD}
\end{equation}
Note that the value of $S_{T < T_*}$ is again
cumulated from the region $T \sim T_*$. The
crucial point is that the contribution to the action
$S(A)$ from the tail of the instanton (or large time
$T < T_*$) is finite, i.e. the integral (\ref{SD})
converges (the addition to the action from interval $T
< T_B$ is negligible). Also this contribution is not
dominant, i.e. it is not much greater than the
contribution of the order $|A|^{3/2}$ from small times
($T > T_*$). It means that in our case the instanton
is localized enough in time. Its long-time dynamics
does not destroy the fact that it is the main
fluctuation determining the statistics of large
negative gradients.

\begin{figure}
  \begin{center}
    \includegraphics[width=3in]{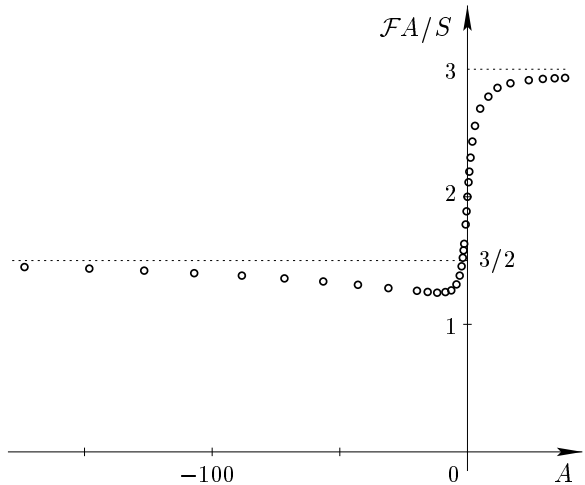}
  \end{center}
  \caption{${\cal F}A/S = d(\ln S)/d(\ln A)$ (see
    (\protect\ref{FSA})) as a function of gradient
    $A$.}
  \label{fig10}
\end{figure}

At the Fig.~\ref{fig10} the function
$d(\ln S)/d(\ln A) = {\cal F}A/S$ that was obtained
from numerical calculations is shown. We used
different grid parameters during calculations for
instanton structure and for this figure. Here we
used $\tau_{\rm min} = -4$ with boundary condition
${\tilde U} (\xi,\tau_{\rm min}) = \chi_2 {\tilde P}
(\xi,\tau_{\rm min})$. This boundary condition was
used as initial condition for ${\tilde U}$ during
iterations. It turned out, that e.g. for ${\cal F}
= -2$ we get the following values of ${\cal F}A/S$
and $A$ with different value of $\tau_{\rm min}$:
\begin{center}
  \begin{tabular}{|c|c|c|}
      \hline
    $\tau_{\rm min}$ & ${\cal F}A/S$ & $A$ \\
      \hline
    $-4$ & $1.441$ & $-148.1$ \\
      \hline
    $-30$ & $1.437$ & $-102.6$ \\
      \hline  
  \end{tabular}
\end{center}
Although, when calculating with $\tau_{\rm min} =
-30$, the condition ${\tilde U}(\xi,-4) = \chi_2
{\tilde P}(\xi,-4)$ holds within 15\% (as one can see
from Fig.~\ref{fig4}) we prefer to use the grid
shorter in time ($\tau_{\rm min} = -4$) to have
smaller time step. The value of $A$ strongly depends
on time step. It was observed in numerical experiment.
Such sensitivity is characteristic of Burgers equation
also. Although the calculations with small $|\tau_{\rm
min}|$ give us worse accuracy at the tail of the
instanton, the smallness of the time step allows us
accurately describe the main part of the instanton
where the nonlinearity level is high.

One can see the cubic asymptotics $S \propto A^3$ at
$A>0$. The instanton structure for $A > 0$ that was
described in \cite{preGM96,prlBFKL97} was confirmed by
our numerical calculations. The case $A<0$
corresponding to the PDF's left tail is more
complicated. The function ${\cal F}A/S$ has minimal
value at $A \simeq -12$. At further decrease of $A$ it
starts to grow and finally tends to the value $3/2$.
In this case the coefficient $S/|A|^{3/2}$ is small.

\section{Conclusion}

We have examined the remote left tail of the velocity
gradients PDF ${\cal P}(u_x)$ in Burgers forced
turbulence. The possibility of direct numerical
solving of instanton equations by iterations is
demonstrated. Numerical calculations and the analysis
of the instanton behavior at the time large compared
with its lifetime $t_* \sim 1/\sqrt{\nu |u_x|}$ with
the solution at small time from \cite{prlBFKL97} show
that $\ln {\cal P}(u_x) \propto -|u_x|^{3/2}$.

\acknowledgments

We are grateful to G.E.~Falkovich, A.V.~Fouxon,
I.V.~Kolokolov, V.V.~Lebedev and E.V.~Podivilov for
useful discussions. This work was partially supported
by Russian Foundation for Basic Research (gr.
98-02-17814), by INTAS (M.S., gr. 96-0457) within the
program of International Center for Fundamental
Physics in Moscow, by the grants of Minerva
Foundation, Germany and Mitchell Research Fund (M.S.).

\end{multicols}

\end{document}